\documentstyle[12pt]{article}
\def\thefootnote{\fnsymbol{footnote}}

\newcommand{\spz}{\hspace{0.7cm}}

\newcommand{\nn}{\nonumber}

\newcommand{\ri}{\right}
\newcommand{\lf}{\left}
\newcommand{\th}{\theta}

\newcommand{\eq}{\begin{equation}}
\newcommand{\en}{\end{equation}}
\newcommand{\bea}{\begin{eqnarray}}
\newcommand{\eea}{\end{eqnarray}}
\newcommand{\acc}{\\[3mm]}
\newcommand{\ba}{\begin{array}}
\newcommand{\ea}{\end{array}}
\newcommand{\ds}{\displaystyle}
\newcommand{\ZZ}{\hbox{{\rm Z{\hbox to 3pt{\hss\rm Z}}}}}
\newcommand{\Y}{\Upsilon}
\newcommand{\CY}{{\cal{Y}}}
\newcommand{\CX}{{\cal{X}}}
\newcommand{\CZ}{{\cal{Z}}}
\newcommand{\BCX}{\bar{{\cal{X}}}}
\def\bx{\bar{x}}
\newcommand{\virg}{\spz,}
\newcommand{\pu}{\spz.}

\newcommand{\NP}[1]{Nucl.\ Phys.\ {\bf #1}}
\newcommand{\PL}[1]{Phys.\ Lett.\ {\bf #1}}

\newcommand{\CMP}[1]{Comm.\ Math.\ Phys.\ {\bf #1}}
\newcommand{\PR}[1]{Phys.\ Rev.\ {\bf #1}}
\newcommand{\PRL}[1]{Phys.\ Rev.\ Lett.\ {\bf #1}}
\newcommand{\MPL}[1]{Mod.\ Phys.\ Lett.\ {\bf #1}}
\newcommand{\IJMP}[1]{Int.\ J.\ Mod.\ Phys.\ {\bf #1}}

\begin{document}
\begin{titlepage}
\vskip 0.5cm
\begin{flushright}
DFTT-39/94 \\
November, 1994\\
hep-th/9411203
\end{flushright}
\vskip0.5cm
\begin{center}
{\large {\bf
ADE functional  dilogarithm identities}\\
{\bf
and}\\
{\bf
integrable models}}\\
\end{center}
\vskip 0.6cm
\centerline{F. Gliozzi$^a$ and R.Tateo$^{a\,b}$}
\vskip 0.6cm
\centerline{\sl  $^a$ Dipartimento di Fisica
Teorica dell'Universit\`a di Torino}
\centerline{\sl Istituto Nazionale di Fisica Nucleare, Sezione di Torino
\footnote{e-mail: gliozzi@to.infn.it}}
\centerline{\sl via P.Giuria 1, I--10125 Torino, Italy}
\vskip .2 cm
\centerline{\sl $^b$ Department of Mathematics, University of
Durham\footnote{e-mail:
tateo@to.infn.it, roberto.tateo@Durham.ac.uk}}
\centerline{\sl   Durham, DH1 3LE England}
\vskip.6cm
\begin{abstract}
\vskip0.2cm
We describe a new  infinite family  of multi-parameter
functional equations for the Rogers dilogarithm, generalizing
Abel's and Euler's formulas. They are suggested by the Thermodynamic
Bethe Ansatz
approach to the renormalization group flow of 2D integrable, ADE-related
quantum field theories. The  known
sum rules for the central charge of critical fixed points can be
obtained as special cases of these. We conjecture that similar
functional identities  can be constructed for any  rational
integrable quantum field theory with factorized S-matrix and support
it with extensive numerical checks.
\end{abstract}
\end{titlepage}

\setcounter{footnote}{0}
\def\thefootnote{\arabic{footnote}}

\section{Introduction}

The Rogers' dilogarithm function~\cite{lewin,kir0} has many
intriguing properties with important implications in many branches of
mathematics and physics.
Recently, in the context of integrable two-dimensional  quantum field
theories, it has been observed that the effective central charge
$\tilde c$ of the conformal field theory describing ultraviolet (UV)
or infrared (IR) fixed points of the renormalization group flow can be
expressed through the Rogers dilogarithm $L(x)$ evaluated at certain
algebraic numbers $x_i$ leading to sum-rules of the
type~\cite{bkr}-\cite{nah}
\eq
\sum_i L(x_i)=\frac{\pi^2}6\tilde c
\label{sum}
\en
In this way it has been found, besides known identities, a large class of
new sum-rules. Many of them are, strictly speaking, only conjectures,
even if they are strongly supported by extensive numerical checks.
Some of them have also been proven using character identities of
Rogers-Ramanujan type~\cite{nah} or other analytic means~\cite{kir0}.

The aim of this paper is to show that these identities are special
cases of new functional equations, written in Eq.s (\ref{re}), (\ref{su})
and (\ref{hh}), where the $x_i$ are rational
functions of a set of variables.
In the ADE-related models the number of independent variables
can be expressed in terms of the rank of the corresponding Lie algebras.

Before attempting to build up these new identities it is helpful to
briefly describe the framework where these identities arise and their
physical motivations. A formal proof is postponed in the next section.

The RG evolution of a two-dimensional integrable theory is described,
in the Thermodynamic Bethe Ansatz approach (TBA), by the ground state
energy $E(R)$ of the system on
an infinitely long cylinder of radius $R$. The equations known (or
conjectured) to give $E(R)$ are of the form
\eq
E(R)=-\frac1{2\pi}\sum_{a=0}^{a=N}\int_{-\infty}^\infty d\theta\;
\nu_a(\theta)\log\left(1+Y_a(\theta)\right)~~~,
\label{en}
\en
where the $Y_a(\theta)$ are $R$ dependent functions determined by a set
of coupled integral equations known as TBA equations; the $\nu_a(\th)$
are suitable elementary functions ( usually
$\nu_a(\th)=c_a\exp(\pm\theta)$ or $\nu_a=c_a\cosh(\theta))$
describing the asymptotic behaviour of the
solutions: $Y_a(\theta)\to \nu_a(\theta)$ for $\theta\to\pm\infty$.

A crucial observation of ref.~\cite{ys} was that any  solution
$\{Y_a(\theta)\}$ of the TBA equations satisfies a set of simple
functional algebraic equations, called theY-system. Conversely it is easy
to show that a set of entire functions satisfying the Y-system with  a
suitable asymptotic behaviour is a solution of the TBA equations, thus
the Y-system encodes all the dynamical properties of the model.

In~\cite{km1,km2,ys,ku,rtv}  a large class of TBA systems
classified according to the ADET  Dynkin's diagrams was proposed to
describe integrable perturbed coset theories~\cite{halp,halp1,gko}
like the ${G_k \times G_l \over G_{k+l}}$ model   with $G$ a  group
associated to  one of the $ADE$  simply-laced algebra, the
${G_k \over U(1)^r}$ generalized parafermionic models,
the ${SU(n)_k \over SO(n)_{2k}}$ , ${SO(2n)_k \over SO(n)_k SO(n)_k}$ ,
${ (E_6)_k \over Sp(8)_k }$ , ${ (E_7)_k \over SU(8)_k }$ ,
${ (E_8)_k \over SO(16)_k }$ perturbed theories, particular points of
the fractional super-sine-Gordon models etc.~\cite{faza,fen}.
The Y-systems  associated to all these models can  be written in terms
of an ordered pair $G\times H$ of $ADET$ Dynkin diagrams in the
following form
\eq
Y_a^b\lf(\th+ \imath { \pi \over \tilde{g}}\ri)
Y_a^b\lf(\th- \imath {\pi \over \tilde{g}}\ri)=
\prod_{c=1}^{r_G} \lf( 1+Y_c^b(\th) \ri)^{G_{a\,c}}
\prod_{d=1}^{r_H} \lf( 1+{1 \over Y_a^d(\th)} \ri)^{-H_{b\,d}}
\label{yy}
\en
where $G_{a\,c}$ and $H_{b\,d}$ are the adjacency matrices of
the corresponding  $ADET$ Dynkin diagram, $\tilde{g}$ is the dual
Coxeter number of $G$,  $r_G$ and $r_H$
are the ranks of the corresponding  algebras.
The functional equations  (\ref{yy}) are  universal in the sense that
using different IR boundary conditions, they  describe
different theories or different regimes  of the same theory.

These algebraic equations may be used as recursion relations: starting
from a set of arbitrary values assigned to the functions $Y_a^b$ at
$\th$ and $\th- \imath { \pi \over \tilde{g}}$ it is possible to
evaluate the same functions at $\th+ \imath { \pi \over \tilde{g}}$.

The main property of these recursion relations is that they generate
{\sl periodic} functions. More precisely, denoting with $\tilde{h}$ the
dual Coxeter number of $H$, for whatever choice of the
initial values of $Y_a^b$ one can verify by direct successive
substitutions or, in the high rank cases, by numerical computations
that~\cite{ys,rqt}
\eq
Y_a^b\lf(\th+ \imath  \pi{\tilde{h} +\tilde{g}
 \over \tilde{g}} \ri) = Y_{\bar a}^{\bar b}(\th ) \virg
\label{xx}
\en
where $\bar a$ and $\bar b$ denote the nodes of the Dynkin diagram
conjugate to $a$ and $b$; for instance, in the $A_n$ diagrams one has
$\bar a=n+1-a$.
This periodicity has many important consequences and is in  relation
with  the conformal dimension of the perturbing operator in the UV
region~\cite{ys}.

Near the fixed points ($R\to0$ or $R\to \infty$) $E(R)$ tends to the
asymptotic form~\cite{bcna}
\eq
E(R)\sim-\frac{\pi\tilde{c}}{6R}~~~,
\label{cas}
\en
where $\tilde c$ is the aforesaid central charge, and the
$Y$'s approach constants $Y_a^b(\th)\to y_a^b$ which can be determined,
for consistency, by the algebraic equations obtained from the Y-system
dropping out the $\th$ dependence.
Combining Eq.s (\ref{en}) and (\ref{cas}) with the $\th$ independent
form of the Y-system, one gets~\cite{kir0,km1,ku,rtv}
\eq
\sum_{a=1}^{r_G} \sum_{b=1}^{r_H} L\lf( \frac{y_a^b}{1 + y_a^b }\ri)
= {\pi^2  \over 6} { r_G  r_H  \tilde{g} \over \tilde{h} +\tilde{g}}
\label{di1}~~~.
\en
This equation represents a wide class of identities of the type
(\ref{sum}). In the next section we shall describe a functional
extension of them.

We conclude this section outlining the physical motivations
suggesting the existence of such a generalization. It can be understood
in a heuristic way in two steps.

First, note  that the effective central charge $\tilde c$ of the Casimir
energy of Eq.(\ref{cas}) measures  in
some sense the number of massless degrees of freedom of the
theory~\cite{km1,bn,cth}. As a consequence, if such a theory flows to a
non-trivial IR limit, $\tilde c_{IR}$ is a RG invariant quantity,
because these degrees of freedom  are not washed away by the process of
renormalization.

In order to take a step forward,  note that this RG invariant should
be expressed in terms of the $Y_a(\th)$'s functions, as  they encode
the dynamical properties of the system everywhere in $R$.
Note also that the RG trajectory connecting the UV
to the IR fixed points is by no means unique. Instead of connecting
these varying the radius of a regular cylinder, we
may interpolate them with any one-parameter family of surfaces
describing a homotopic deformation of the UV cylinder into the IR one.
It is conceivable that some of these trajectories correspond
to integrable flows. We assume, as it seems  reasonable,  that these
are  described by the same Y-system with a modified
asymptotic behaviour, which becomes a function of the trajectory.
On the other hand, the RG invariant $\tilde c_{IR}$ should not depend
on the choice of the trajectory, of course. We are then faced to an
apparent dilemma:  deforming  a RG trajectory produces a
modification of the $Y_a(\th)$'s keeping invariant $\tilde c_{IR}$,
which is a function of them. A possible way out is to guess that
$\tilde c_{IR}$ is given by a sort of topological invariant of the
Y-system: its value should be determined by the fact that the
$Y_a(\th)$'s satisfy the recursion relations of Eq.(\ref{yy}) and not
by the analytic properties of them.

Then one expects a generalization of Eq.(\ref{di1})
where the constants $y_a$ are suitably replaced by an {\sl arbitrary}
set of solutions $Y_a(\th)$  of the Y-system.

\section{New functional dilogarithm identities}

The Rogers dilogarithm $L(x)$ with $0\le x\le1$ is the unique function
that is three times differentiable and satisfies the following five term
relationship  known also as the Abel functional equation (see for
instance ref.~\cite{kir0})
\eq
L(x)+L(1-xy)+L(y) +L \lf( \frac{1-y}{1-xy}\ri) +
L\lf(\frac{1-x}{1-xy}\ri)=3\,L(1)~~~,
\label{abel}
\en
with $0\le x,y\le1$ and the  normalization $L(1)=\pi^2/6$.

To begin with, we observe that
this relation has an hidden pentagonal symmetry. Indeed, denoting the
five arguments of $L$ in the {\sl same} order as they appear in
Eq.(\ref{abel}) by $a_n$ with $n=0,1\dots4$, one can verify at once that
they satisfy for arbitrary values of $x$ and $y$ the following
recursion relation
\eq
a_{n-1}a_{n+1}=1-a_n~~~,
\en
which has an intriguing ${\ZZ}_5$ symmetry:
\eq
a_{n+5}=a_n~~~.
\label{z5}
\en
One can easily recognize a similarity between such a {\sl
periodic} recursion  relation and the Y-systems of Eq.(\ref{yy}).
Such a similarity can be made even more strict by reshuffling
the variables in the following way: taking\footnote{This permutation
is an outer automorphism $n\to n'$ of the underlying $\ZZ_5$ symmetry
generated by the congruence $n'\equiv 3n \;{\rm mod}(5)$}
$b_1=a_3\,,\,b_2=a_1\,,\,b_3=a_4\,,\,b_4=a_2\,,\,b_5=a_5\,$ and putting
$b_n=\frac{\CY_n}{1+\CY_n}$ we get the following set of {\sl periodic}
recursion relations
\eq
\CY_{n-1}\CY_{n+1}=1+\CY_n~~~,
\label{A2}
\en
where one gets again $\CY_{n+5}=\CY_n~$. Actually Eq.(\ref{A2}) is a
slightly disguised form of the $A_1\times A_2$ Y-system describing the
RG flow of the tricritical Ising fixed point to the critical Ising one.
Indeed, putting $i(n)=2+(-1)^n$, comparison with Eq.(\ref{yy}) yields
\eq
\CY_n= 1 / Y_1^{i(n)}\lf( \th+ \imath \frac{n\pi}2 \ri)~~~,
\en
Showing that the five term functional relation may be viewed as a
dilogarithm identity satisfied by an arbitrary solution of the
$A_1\times A_2$ Y-system. Thus one is naturally led to conjecture that
a similar property should  hold also for the other Y-systems. In order
to write explicitly this new family of identities it is useful to
simplify the notation by taking
\eq
\Y_a^b(n)=Y_a^b\left(\th+ \imath \frac{n\pi}{\tilde{g}}\right)~~n=0,1,2,
\dots
\en
Then the recursive relations of the Y-system of Eq.(\ref{yy}) can be
written as
\eq
\Y_a^b(n+1) \Y_a^b(n-1) =
\prod_{c=1}^{r_G} \lf( 1+\Y_c^b(n) \ri)^{G_{a\,c}}
\prod_{d=1}^{r_H} \lf( 1+{1 \over \Y_a^d(n)} \ri)^{-H_{b\,d}}~~~.
\label{re}
\en
According to our conjecture, any solution of such recursive equations
must satisfy the following  identity
\eq
\sum_{a=1}^{r_G} \sum_{b=1}^{r_H}
\sum_{n=0}^{\tilde{h}+\tilde{g}-1} H( \Y_a^b(n)) =
r_G r_H \tilde{g} ~~~,
\label{su}
\en
where we introduced for our  notational convenience  the function
\eq
H(x)={ 6 \over \pi^2 } L\lf({x\over 1+x} \ri) \pu
\label{hh}
\en
Before attempting to prove these new identities it is useful to note
that  they are multi-parameter generalizations of known identities,
indeed putting the $n$-independent solution  of Eq.({\ref{re}) into
Eq.(\ref{su}) one  gets at once Eq.({\ref{di1}). Notice also that the
Y-systems belonging to the subset of those lacking of the tadpole
diagrams of the type $T_n$ can be split in a pair of independent
algebraic systems. As a consequence the identity (\ref{su}) is
split  accordingly in a pair of equivalent identities, of course.
For instance in a $A_l\times A_m$ Y-system we can constrain the triple
sum of Eq.(\ref{su}) to the subset in which $a+b+n$ is an even number;
consistently we have to divide by two the right-hand side of the
equation.
In particular the simplest identity, associated to the
$A_1\times A_1$ system, gives
\eq
H(x)+H \lf({1\over x}\ri)=1~~~,
\label{ab}
\en
which is nothing but the Euler identity $L(y)+L(1-y)=L(1)$.
The Abel equation can be written within these notations as
\eq
\sum_{n=0}^4 H(\CY_n)=3~~~.
\label{ab1}
\en
The simplest new identity of the type (\ref{su}) is associated to the
Y-system $A_1\times A_3$. In order to prove it, let
us indicate for convenience the $\Y(n)$ associated to the central node
of $A_3$ by $\CZ_n$ and those associated to the other two by $\CX_n$ and
$\bar{\CX}_n$. Then we can write the following periodic recursive
equations with underlying $\ZZ_6$ symmetry
\bea
\CX_n\CX_{n+2}=&1+\CZ_{n+1}\nn\\
\bar{\CX}_n\bar{\CX}_{n+2}=&1+\CZ_{n+1}\nn\\
\CZ_{n-1}\CZ_{n+1}=&(1+\CX_{n})(1+\bar{\CX}_{n})~~~,
\label{A3}
\eea
where the mentioned splitting between odd and even variables is made
evident. One can also verify at once that $\CX_n=\BCX_{n+6}$
and $\CZ_n=\CZ_{n+6}$

The corresponding identity can be written as
\eq
\sum_{m=0}^2\left(H(\CX_{2m})+H(\CZ_{2m+1})+H(\BCX_{2m})\right)=6~~~,
\label{id}
\en
or in more explicit way, solving the recursion relations in terms of the
three free parameters $x=\CX_0$, $y=\CZ_1$ and $\bx=\BCX_0$, as
\eq
\ba{c}
\ds{  H\lf( x \ri) +   H\lf({1 + y \over x } \ri) +
    H\lf({1 + x + y+\bx+\bx x\over y\bx}\ri) + } \acc
\ds{ H(y)+H\lf({(1 + x + y) (1 + y + \bar{x})  \over
           x y \bar{x}  }\ri) +
H\lf({(1 + x)(1 + \bar{x}) \over y}\ri) + } \acc
\ds{  H\lf( \bx \ri) +   H\lf({1 + y \over \bx } \ri) +
    H\lf({1 + x + y+\bx+\bx x\over yx}\ri) = 6\pu } \acc
\label{gg}
\ea
\en
Eq.(\ref{A3}) tells us that the three triplets
$\{\BCX_{0},\CZ_{1},\BCX_{2}\}\,$ , $\{\CX_{2},\CZ_{3},\CX_{4}\}\,$ and
$\{\BCX_{4},\CZ_{5},\CX_{0}\}$, can be considered as the first three
terms $\CY_0,\CY_1,\CY_2$ of Eq.(\ref{A2}), so we can use
Eq.(\ref{ab1}) to replace  the sum of the nine terms of the left-hand
side of Eq.(\ref{id}) to the following sum of six $H$ terms
\[
9 -H\lf( { y \bar{x} \over 1+y+\bar{x}} \ri)
-H\lf( {x (1+\bar{x}) \over 1+y + \bar{x} }\ri)
-H\lf( { 1+y+\bar{x} \over x (1+\bar{x}) } \ri)
\]
\[
-H\lf( { y \over 1 + \bar{x} }\ri)
-H\lf( { 1+ \bar{x} \over y} \ri)
-H\lf( {1+y + \bar{x}  \over y \bar{x}}\ri)
\]
which can be split in three pairs of the form $H(z)+H(\frac1z)$, so
using Eq.(\ref{ab}) we get that the LHS is equal to 6 as it was stated.

Similarly one can prove these identities for other low rank
systems. As the rank increases the direct algebraic manipulations become
rather involved and we not succeeded in finding a general recursive proof.
For higher rank algebras, we did extensive numerical checks.

We conclude with two important observations.

There is a class of  integrable models where the TBA equations or the
corresponding Y-system is not apparently related to the Dynkin diagrams.
This class should include the sine-Gordon model at rational points
and  its quantum reduced models as well all the reduced models of the
affine Toda field theories at imaginary coupling constant~\cite{smi,hol}.
The physical motivations described in \S 1 suggest that also in these
cases identities of the kind (\ref{su}) should hold.
In order to support this conjecture we used  the  TBA equations
proposed in~\cite{ku,ta,fen} and  others associated to models with
purely elastic S-matrix (see for instance~\cite{mar,mu}) and we
verified  the corresponding dilogarithm identity for the associated Y-
system by extensive numerical checks.

Finally, note that the periodicity of the Y-system, although it plays a
fundamental role in many properties  of the integrable models, it has not
been fully understood. Actually it has not been proven in a general
way, but it has been verified by direct algebraic
manipulations only for algebras of low rank. Yet the validity of
the Rogers dilogarithm identities discussed above are strictly
related to it: if the $\Y$ variables were not periodic Eq.(\ref{su})
would be meaningless. It would be interesting to understand this
property in a more general way.

 \vskip .6cm
{\bf Acknowledgements} --
R.T. would like to thank P.Dorey for useful discussions and the
Mathematics Department of Durham University for the kind hospitality.

\end{document}